\documentclass[%
reprint,
amsmath,amssymb,
prb,
floatfix,
]{revtex4-2}

\usepackage{graphicx}
\usepackage{dcolumn}
\usepackage{bm}
\usepackage{color}

\begin{document}

\title{Pair binding and enhancement of pairing correlations in asymmetric Hubbard ladders}
\author{Anas Abdelwahab}
\author{G\"{o}kmen Polat}
\author{Eric Jeckelmann}
\affiliation{Leibniz Universit\"{a}t Hannover, Institute of Theoretical Physics,  Appelstr.~2, 30167 Hannover, Germany}

\date{\today}

\begin{abstract}
Asymmetric two-leg Hubbard ladders with different on-site interactions $U_y$ and hoppings $t_y$ on each leg are investigated using the density matrix renormalization group method and exact diagonalizations. 
The pairing found in symmetric  ladders is robust against the introduction of the leg asymmetry.
When studying pairing, one-band Hubbard ladder models are better described as one-dimensional correlated two-band models than as sublattices of higher dimensional systems.
The asymmetric Hubbard ladder provides us with a simple model for studying pairing in the crossover regime between charge transfer and 
Mott insulators.
\end{abstract}

\maketitle

\section{Introduction.} 
Correlated electron models on ladder lattices have been studied for almost 30 years with the aim of shedding light on the origin of unconventional superconductivity in layered and ladder-like cuprate compounds \cite{Dagotto1994,Scalapino2012,ric93,Maekawa1996,Uehara1996}.
The key idea is that the one-dimensional (1D) ladder-like sublattice is less difficult to study but retains the essential properties
of the two-dimensional (2D) lattice.
This approach is still widely used because investigations of truly 2D models for 
correlated electrons remain extremely challenging~\cite{LeBlanc2015}.
Many studies have focused on lightly doped two-leg ladders with equivalent legs. 
They have shown that the low-energy properties of one-band Hubbard ladders are consistent with a Luther-Emery 
phase~\cite{Luther1974}
with d-wave-like (quasi-long-range) pairing correlations \cite{noa94,bal96,noa97,dol15,Gannot2020},
which can be seen as the 1D precursor of a d-wave superconducting phase in higher dimensions. 

Investigations of the more generic but also more challenging three-band Hubbard model~\cite{eme87}
 on ladders have been rare
and less conclusive~\cite{Jeckelmann1998c,Nishimoto2002b,Nishimoto2009,son21}. 
A recent study did not found a Luther-Emery phase
in that model with realistic parameters 
for cuprates~\cite{son21}. 
It is also noticeable that recent studies suggest a completely opposite situation in two dimensions:
the one-band Hubbard model is likely not superconducting~\cite{qin20} while the three-band Hubbard model
could describe the superconducting phase of layered cuprates~\cite{Mai2021}.
These contradictions call into questions the adequacy of correlated ladder models 
for studying unconventional superconductivity in cuprates.

\begin{figure}
\includegraphics[width=0.49\textwidth]{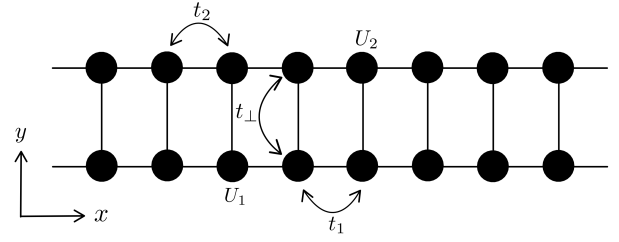}
\caption{\label{fig01} Sketch of the asymmetric Hubbard ladder
  described by the Hamiltonian~(\ref{eq:asymmetric_Ladder}) with
  different onsite repulsions $U_y$ and 
   hoppings $t_y$  on each leg ($y=1,2$) and interchain hopping
  $t_\perp$. 
}
\end{figure}

Here we report on our study of 
an asymmetric two-leg Hubbard ladder with different on-site interactions $U_y$ and hoppings $t_y$
on each leg~\cite{yos09}, see Fig.~\ref{fig01}. This model can also be seen as a two-band Hubbard chain~\cite{AlHassanieh2009}.
Thus it can describe both the doping
of a Mott insulator like the one-band Hubbard model and the doping of a charge-transfer insulator
like the three-band Hubbard model, but it is simpler than the latter and includes the former as a special case.

Our results show that the pairing found in symmetric Hubbard ladders is robust against the introduction of a leg asymmetry.
Thus when studying pairing properties, one-band ladder models are better seen as 1D two-band correlated models than as sublattices of higher dimensional lattices.
 We conclude that  asymmetric ladder models provide us with an interesting opportunity to study pairing upon doping in systems lying in between charge transfer and Mott insulators.
 
\section{Model and methods.} 
The Hamiltonian of the asymmetric two-leg Hubbard ladder shown in Fig.~\ref{fig01} is
\begin{eqnarray}\label{eq:asymmetric_Ladder}
H=
&-&\sum_{x,y,\sigma} t_{y} \left ( c_{x+1,y,\sigma}^{\dagger}c_{x,y,\sigma}^{\phantom{\dagger}} 
+ c_{x,y,\sigma}^{\dagger}c_{x+1,y,\sigma}^{\phantom{\dagger}} \right )  \nonumber \\
&-&t_{\perp}\sum_{x,\sigma} \left ( c_{x,2,\sigma}^{\dagger} c_{x,1,\sigma}^{\phantom{\dagger}}  +
c_{x,1,\sigma}^{\dagger}c_{x,2,\sigma}^{\phantom{\dagger}} \right ) \nonumber \\
&+& \sum_{x,y} U_{y} \left ( n_{x,y,\uparrow}-\frac{1}{2} \right ) \left ( n_{x,y,\downarrow}-\frac{1}{2} \right )\,.
\label{eq:hamiltonian}
\end{eqnarray}
where $c_{x,y,\sigma}$($c^{\dagger}_{x,y,\sigma}$) is an annihilation (creation)
operator for an electron with spin $\sigma$ on the site with coordinates $(x,y)$,  $y=1$ (first leg) or $y=2$ (second leg), and the rung index $x$ runs from $1$ to the ladder length $L$.
The electron number operators are denoted $\hat{n}_{x,y,\sigma} = c_{x,y,\sigma}^{\dagger} c_{x,y,\sigma}^{\phantom{\dagger}}$.
The range of the average site density  is $0\leq n \leq 2$ with $n=1$ corresponding to a half-filled ladder.
As the Hamiltonian is symmetric under a particle-hole transformation,
we discuss only electron doping, i.e. $n \geq 1$. The particle-hole symmetry
can be removed to mimic the asymmetry in superconducting cuprates using on-site potentials that differ from
$-U_y/2$ but we have not explored this generalization yet. 

The Hamiltonian~(\ref{eq:hamiltonian}) depends on 5 parameters 
($t_{\perp},t_y, U_y, y=1,2$).
As both legs and their relative phase can be exchanged without changing the model properties, 
we can restrict ourselves to the case $t_{\perp} \geq 0$ and $\Delta U = U_1 - U_2 \geq 0$.
The single-particle spectrum of the noninteracting asymmetric two-leg ladder ($U_y=0$) is made
of two bands of equal width $4t_{\parallel}$ with the average leg hopping $t_{\parallel}=\vert t_1+t_2 \vert /2$.
Thus all energies presented here are given in units of $t_{\parallel}$. Moreover, it is convenient to introduce
$\Delta t=t_1-t_2$ as a measure of the asymmetry of intra-leg hoppings. 
In the dimer limit ($t_{\perp} \gg \vert t_y \vert$)
the eigenstates are more conveniently expressed with the average interaction $\bar{U}=(U_1+U_2)/2$
and the difference $\Delta U = U_1 - U_2$.
However, we find that the pairing properties
depend essentially on the ratio 
$u_y=U_y/t_y$. Thus we will use both parametrizations in the discussion of our results.

Here we focus on two quantities, the pair binding energy and pair correlation functions.
Let $E_{0}\left( N_{\uparrow}, N_{\downarrow} \right)$ be the ground-state energy of the ladder system with $N_{\sigma}$ electrons of spin $\sigma$.
The pair binding energy (PBE) for two added ($+$) or two removed ($-$) electrons is given by
\begin{eqnarray}\label{eq:EpbPlusMinusDMRG}
    E^{\pm}_{pb} &=& 2E_0\left( N_{\uparrow} \pm 1,N_{\downarrow} \right) - E_0\left( N_{\uparrow} \pm 1,N_{\downarrow} \pm 1 \right) \nonumber \\
    &-& E_0 \left( N_{\uparrow},N_{\downarrow} \right) .
\end{eqnarray}
The single-particle gap is defined as
\begin{eqnarray}\label{eq:EpDMRG}
    E_{p} &=& E_0\left( N_{\uparrow} + 1,N_{\downarrow} \right) + E_0\left( N_{\uparrow} - 1,N_{\downarrow} \right) \nonumber \\
    &-& 2E_0 \left( N_{\uparrow},N_{\downarrow} \right)
\end{eqnarray}
and the charge gap is defined as
\begin{eqnarray}\label{eq:EcDMRG}
    E_{c} &=& \frac{1}{2} \left[ E_0\left( N_{\uparrow} + 1,N_{\downarrow} +1 \right) + E_0\left( N_{\uparrow} - 1,N_{\downarrow} -1 \right) \right. \nonumber \\
    &-& \left. 2E_0 \left( N_{\uparrow},N_{\downarrow} \right) \right ] .
\end{eqnarray}
Moreover, the spin gap is defined as
\begin{equation}\label{eq:EsDMRG}
    E_{s} = E_0\left( N_{\uparrow} + 1,N_{\downarrow} - 1 \right)
    - E_0 \left( N_{\uparrow},N_{\downarrow} \right) .
\end{equation}
At half filling the PBE becomes
\begin{equation}\label{eq:EpbHFDMRG}
    E_{pb} = E^{+}_{pb} = E^{-}_{pb} = E_p - E_c .
\end{equation}
In our discussion of the PBE we present results for $E^{-}_{pb}$ only. $E^{+}_{pb}$ give similar results with negligible differences. 

A positive PBE means that the system gains energy when doped electrons build
bound pairs. Thus it is a necessary but not sufficient condition for the 1D precursor 
of a superconducting phase. Better evidence is given by the long-range behavior
of 
correlation functions although this is  more difficult to determine accurately
with numerical methods.
We have investigated general pairing correlation functions  of the form
\begin{equation}\label{eq:correlations}
C = \left \langle c_{x_0,y_0,\uparrow}^{\dagger}c_{x'_0,y'_0,\downarrow}^{\dagger} c_{x',y',\downarrow}^{\phantom{\dagger}}c_{x,y,\uparrow}^{\phantom{\dagger}} \right \rangle
\end{equation}
where electron pairs  are created and annihilated on two pairs of nearest-neighbor sites,  $\{(x_0,y_0)$,$(x'_0,y'_0)\}$
and  $\{(x,y)$,$(x',y')\}$, respectively.
For instance, correlations between a rung pair around the ladder center 
($x_0 \approx L/2$) and a pair on the second leg are given by
\begin{equation}\label{eq:RLcorrelation}
C(x-x_0) = \left \langle c_{x_0,1,\uparrow}^{\dagger}c_{x_0,2,\downarrow}^{\dagger} c_{x,2,\downarrow}^{\phantom{\dagger}}c_{x+1,2,\uparrow}^{\phantom{\dagger}} \right \rangle .
\end{equation}
Rung-rung correlations are obtained with $\{x_0=x'_0, y_0\neq y'_0, x=x', y\neq y'\}$, e.g.
\begin{equation}\label{eq:RRcorrelation}
C_R(x-x_0) = \left \langle c_{x_0,1,\uparrow}^{\dagger}c_{x_0,2,\downarrow}^{\dagger} c_{x,1,\downarrow}^{\phantom{\dagger}}c_{x,2,\uparrow}^{\phantom{\dagger}} \right \rangle .
\end{equation}
while leg-leg correlations correspond
to $\{\vert x_0-x'_0\vert=1, y_0=y'_0, \vert x-x'\vert=1, y=y'\}$, e.g., 
\begin{equation}\label{eq:LLcorrelation}
C_L(x-x_0) = \left \langle c_{x_0,2,\uparrow}^{\dagger}c_{x_0+1,2,\downarrow}^{\dagger} c_{x,2,\downarrow}^{\phantom{\dagger}}c_{x+1,2,\uparrow}^{\phantom{\dagger}} \right \rangle .
\end{equation}
These correlation functions decrease as $(x-x_0)^{-2}$ in the noninteracting limit $U_y=0$. Pairing correlations are said to be enhanced
when they decay more slowly than in a noninteracting system. In 1D  the slowest possible decay is $C(x-x_0) \sim (x-x_0)^{-1}$,
which corresponds to quasi-long-range pairing correlations.

In lightly doped symmetric ladders, i.e. the Hamiltonian~(\ref{eq:hamiltonian}) with $t_1=t_2=t$ and $U_1=U_2=U>0$, one finds a positive PBE accompanied by an enhancement of pairing correlations with a d-wave-like structure at intermediate coupling $U/t$ ~\cite{noa94,bal96,noa97,dol15}. This phase with gapless charge excitations but gapped spin excitations is consistent with a  Luther-Emery phase~\cite{Luther1974}.
Robust pairing was found in a Kondo-Heisenberg model~\cite{AlHassanieh2009}
 that can be obtained in the strong coupling limit $U_y \gg \vert t_y\vert, \vert t_{\perp}\vert$
of the model~(\ref{eq:hamiltonian})
while pair-density-wave (PDW) correlations
were found to be enhanced in another Kondo-Heisenberg model~\cite{ber10}
that corresponds to the limiting case $U_1 \gg\vert t_y\vert, \vert t_{\perp}\vert$ with $U_2=0$. 
In the special case of a single noninteracting leg ($U_2=0$ with $U_1>0$, $t_1=t_2$)
a positive pair binding energy is obtained 
upon doping away from half filling 
for intermediate couplings 
but no enhancement of pairing correlations has been found~\cite{abd15,abd18}.
To the best of our knowledge the intermediate regime with finite but different parameters $U_y$ and $t_y$ 
in both legs has not been investigated yet.

We investigate the asymmetric ladder (\ref{eq:asymmetric_Ladder}) using the density matrix renormalization 
group (DMRG) method~\cite{White1992,White1993,Schollwoeck2005,Jeckelmann2008,Schollwoeck2011} and exact diagonalizations~\cite{Weisse2008}.
DMRG has been extensively and successfully used to study ladder systems for almost three decades~\cite{noa94,Schollwoeck2005}.
In particular, it was employed to investigate pairing effects in various ladders with inequivalent legs~\cite{AlHassanieh2009,ber10,abd15,abd18}, which
correspond to limiting cases of~(\ref{eq:asymmetric_Ladder}).
We employ DMRG to compute the ground-state properties of ladders with open boundary conditions and even number of rungs up to $L=200$. 
We keep up to $m=2048$ density-matrix eigenstates in our calculations yielding discarded weights of the order of $10^{-6}$ or smaller.
We always repeat the calculations for several numbers $m$ and extrapolate the ground-state energy to the limit of vanishing discarded weights in order to estimate the DMRG truncation error~\cite{Bonca2000}.
Additionally, we compute the properties of the ground states and lowest excited states in short ladders with periodic boundary conditions and up to $L=6$ rungs
using the Lanczos diagonalization method~\cite{Weisse2008}.

\section{Results}

\begin{figure}
\includegraphics[width=0.40\textwidth]{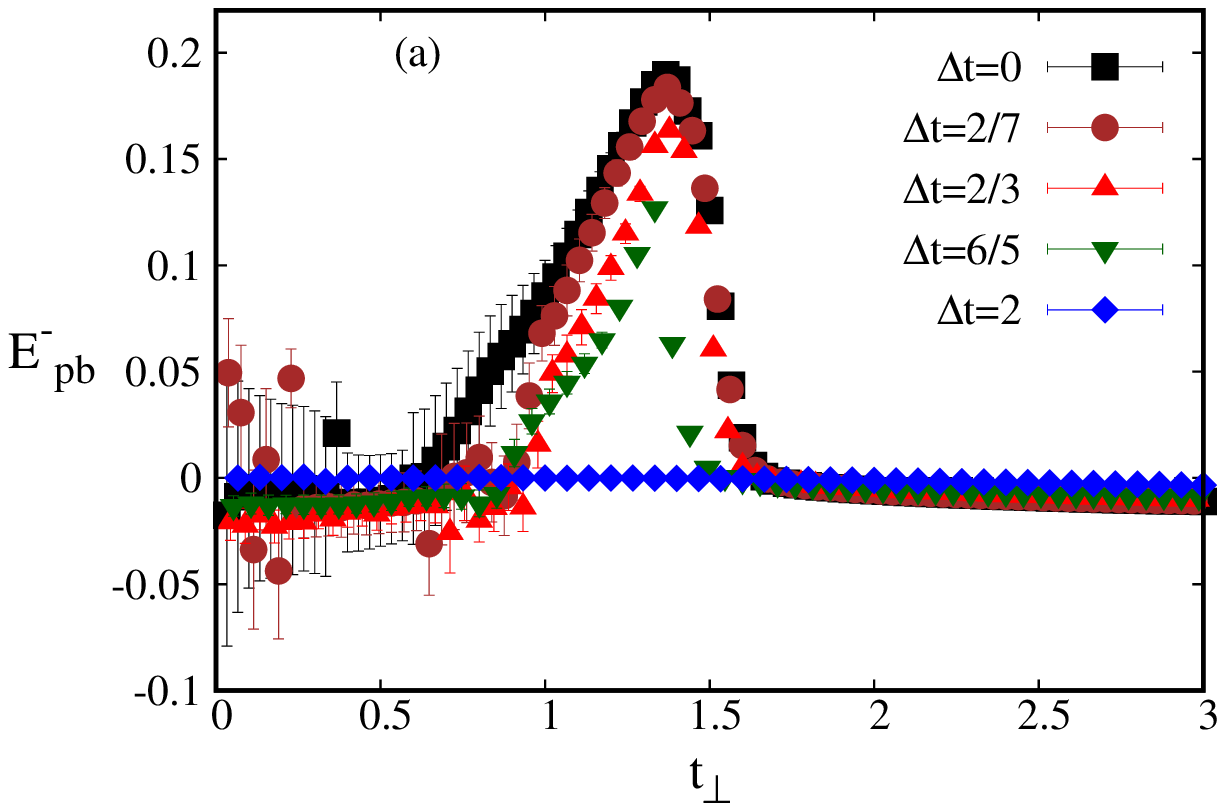}
\includegraphics[width=0.40\textwidth]{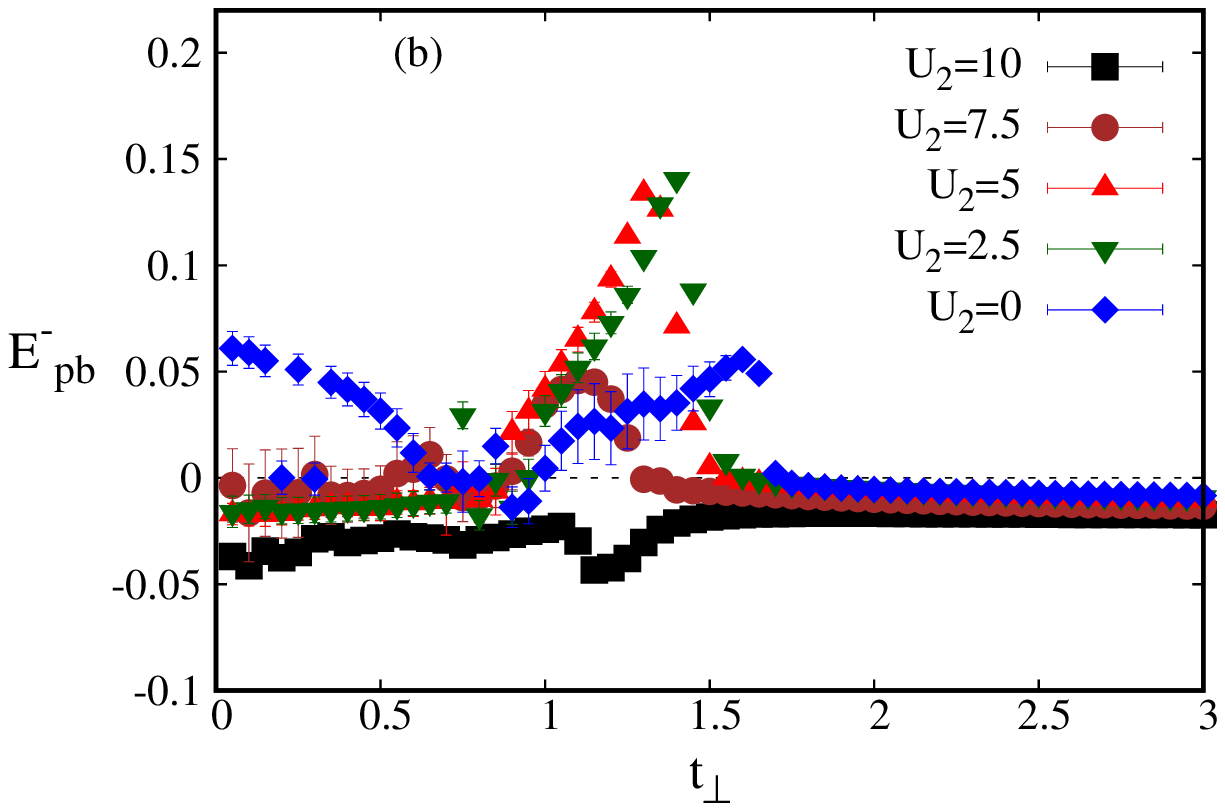}
\includegraphics[width=0.40\textwidth]{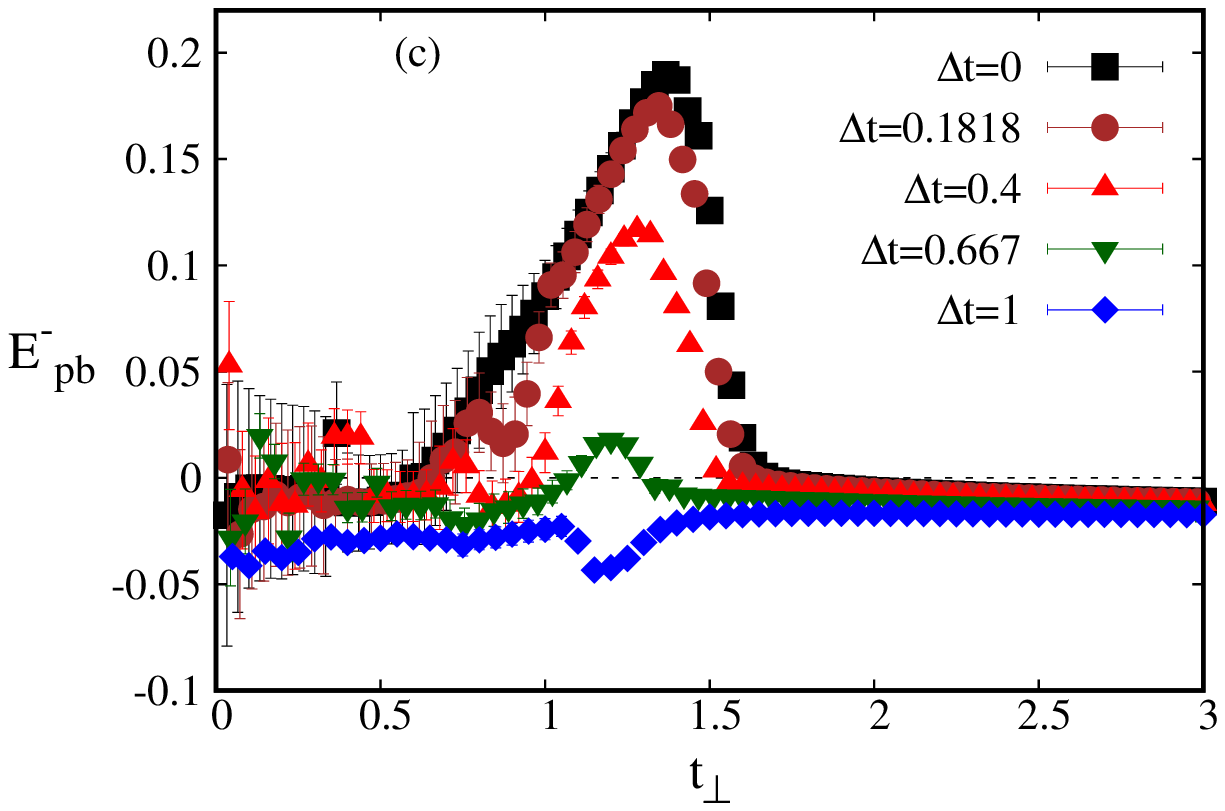}
\caption{\label{fig:U20EpbMinus} 
Pair binding energy $E^{-}_{pb}$ as function of $t_{\perp}$ for ladders with length $L=128$ and average density $n=1.125$.
(a) Ladders with $u_1 = u_2 = 20/3$  and various $\Delta t \geq 0$, see Table~\ref{table:u1_Equal_u2}.
(b) Ladders with $\Delta t =1$, $u_1=20/3$, and various couplings $U_2=u_2 t_2 = u_2/2>0$, see Table~\ref{table:u1_NotEqual_u2}.
(c) Ladders with  $\Delta U=0$ and various $\Delta t$, see Table~\ref{table3}.
Error bars show the size of DMRG truncation errors.
}
\end{figure}

\subsection{Pair binding energy}
We first discuss our findings for the pair binding energy.
Figure~\ref{fig:U20EpbMinus} shows the PBE as function of the interchain hopping $t_{\perp}$ 
for various asymmetries in ladders with $L=128$ rungs and an average density $n=1.125$.
In Fig.~\ref{fig:U20EpbMinus}(a) we quantify and vary  the asymmetry using $\Delta t$ while  $u_1=u_2=20/3$ is kept fixed.
Note that $\Delta U$ also varies. The parameters used in Fig.~\ref{fig:U20EpbMinus}(a) are listed in Table~\ref{table:u1_Equal_u2}.
We observe that the PBE is positive in a narrow range of interchain hopping and exhibits a maximum at $t_{\perp} \approx 1.4 t_{\parallel}$.
See Table~\ref{table:u1_Equal_u2} for the maximum values and positions.
Range and maximum are the largest for the symmetric ladder ($\Delta t = \Delta U =0$) but both
decrease slowly with increasing asymmetry $\Delta t$.
For the largest asymmetry $(\Delta t=2 \Rightarrow t_2=U_2=0)$, however, the PBE vanishes for all interchain hoppings.
We did not investigate whether the region of positive PBE disappears exactly at $\Delta t=2$ or for a lower value.

\begin{table}
\centering
\begin{tabular}{|c||c|c|c|c|c|c||c|c|}
 \hline
 & $t_1$ & $t_2$&$\Delta t$  & $U_1$ & $U_2$ & $\Delta U$ & $t_{\perp}$ & $E^{-}_{pb}$ \\
 \hline
 \#1 & 1 & 1 & 0 & 20/3 & 20/3 & 0 & 1.367 & 0.19 \\
 \#2 & 8/7 & 6/7 & 2/7 & 160/21  & 120/21 & 40/21 & 1.371 & 0.184 \\
 \#3 & 4/3 & 2/3 & 2/3 & 80/9 & 40/9 & 40/9 & 1.378 & 0.163 \\
 \#4 & 8/5 & 2/5 & 6/5 & 32/3 & 8/3 & 8 & 1.333 & 0.127 \\
 \#5 & 2 & 0 & 2&40/3 &0  & 40/3 & - & - \\
 \hline
\end{tabular}
\caption{Model parameters and maximum pair binding energies shown in Fig.~\ref{fig:U20EpbMinus}(a). In all cases $u_1=u_2=20/3$.
}
\label{table:u1_Equal_u2}
\end{table}

In Fig.~\ref{fig:U20EpbMinus}(b) we vary the asymmetry using $u_2$ while  $\Delta t = 1$ and $u_1=20/3$ are kept fixed.
Again this means varying $\Delta U$. The parameters used in Fig.~\ref{fig:U20EpbMinus}(b) are listed in Table~\ref{table:u1_NotEqual_u2}.
We observe again that the PBE is positive in a narrow range of the interchain hopping with maxima around  $t_{\perp} \approx 1.4 t_{\parallel}$
when $u_2$ is close to $u_1$. 
Clearly, the PBE maxima and the ranges where $E^{-}_{pb}$ is  positive diminish when the asymmetry between $u_2$ and $u_1$ becomes larger.
See Table~\ref{table:u1_NotEqual_u2} for the maximum values and positions.
Note that the positive PBE
for small $t_{\perp}$ at $U_2=0$ are due to finite size effects (see the discussion below)
and larger DMRG truncation errors in that limit.
We know exactly that the system is a four-component Luttinger liquid with dominant antiferromagnetic spin correlations for $t_{\perp}=0$~\cite{abd15}.

\begin{table}[b]
\centering
\begin{tabular}{ |c|c|c||c|c|  }
 \hline
 $U_2$& $u_2$  & $\Delta U$ & $t_{\perp}$ & $E^{-}_{pb}$ \\
 \hline
 10 & 20 & 0 & - & - \\
 7.5 & 15 & 2.5 & 1.1 & 0.046 \\
 5 & 10 & 5 & 1.35 & 0.126 \\
 2.5 & 5  & 7.5 & 1.4 & 0.141 \\
 0 & 0 & 10 & 1.6 & 0.056 \\
 \hline
\end{tabular}
\caption{Model parameters and maximum pair binding energies shown in Fig.~\ref{fig:U20EpbMinus}(b).  
In all cases $U_1=10$, $t_1=1.5$, and $t_2=0.5$ ($\Rightarrow \Delta t = 1$ and $u_1=20/3$).
}
\label{table:u1_NotEqual_u2}
\end{table}

Similarly, we can vary the asymmetry using $u_2$ while  $\Delta U$ and $u_1=20/3$ are kept fixed.
This implies that $\Delta t$ varies. This is illustrated in Fig.~\ref{fig:U20EpbMinus}(c) for  $\Delta U=0$.
The parameters used in this figure are listed in Table~\ref{table3}.
We again find that the PBE is positive in a narrow range  of interchain hoppings with 
maxima by $t_{\perp} \approx 1.4 t_{\parallel}$ when $u_1 =  u_2 (\Rightarrow \Delta t = 0)$.
Clearly, maxima and ranges diminish gradually when the asymmetry between $u_2$ and $u_1$ becomes larger.
See Table~\ref{table3} for the maximum values and positions.
For the largest asymmetry ($u_2/u_1=3  \Rightarrow \Delta t = 1$) the PBE is negative for all interchain hoppings


\begin{table}
\centering
\begin{tabular}{ |c|c|c|c|c||c|c| }
 \hline
 $t_1$ & $t_2$&$\Delta t$  & $U_1=U_2$ & $u_2$ & $t_{\perp}$ & $E^{-}_{pb}$ \\
 \hline
 1 & 1 & 0  & 20/3  &  20/3 & 1.367 & 0.1896 \\
 12/11 & 10/11 & 2/11 & 80/11  & 8 & 1.345  & 0.175 \\
 6/5 & 4/5 & 2/5 & 8  & 10 & 1.28 & 0.117 \\
 4/3 & 2/3 & 2/3 & 80/9  & 40/3 & 1.2 & 0.018 \\
 3/2 & 1/2 & 1 & 10  & 20 & - & - \\
 \hline
\end{tabular}
\caption{Model parameters and maximum pair binding energies shown in Fig.~\ref{fig:U20EpbMinus}(c).
In all cases $\Delta U=0$ and $u_1=20/3$.
\label{table3}
}
\end{table}

\begin{figure}[b]
\includegraphics[width=0.35\textwidth]{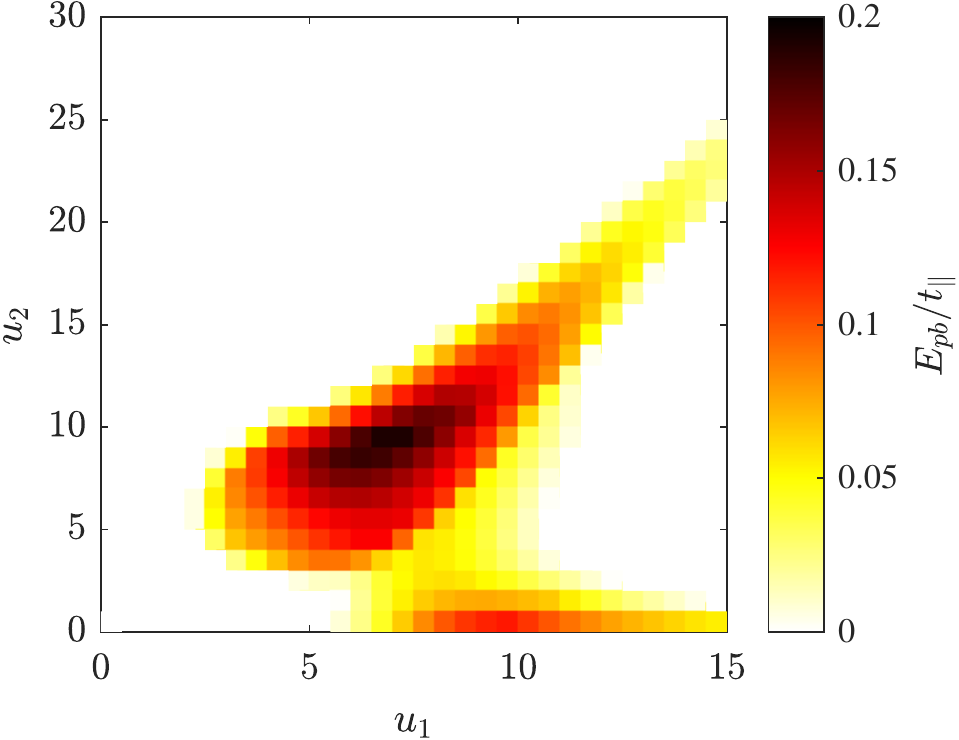}
\caption{\label{fig:exactdiag} 
Pair binding energy $E^{-}_{pb}$ as function of $u_1=U_1/t_1$ and $u_2=U_2/t_2$ for two electrons added to a  half-filled $2 \times 6$ ladder
with  $t_1 = 4/3$, $t_2=2/3$ ($\Rightarrow \Delta t = 2/3$), and $t_{\perp}=4/3$.
}
\end{figure}

Our DMRG results show that positive PBE can be found in very asymmetric ladders as long as
$u_1=U_1/t_1$  and $u_2=U_2/t_2$ remain close. The largest PBE are obtained
with couplings in the intermediate range $u_y=5$ to $10$. We have confirmed these findings using
exact diagonalizations of short asymmetric ladders with periodic boundary conditions
to calculate $E^{-}_{pb}$  systematically as a function of $u_1$ and $u_2$.
For instance, Fig.~\ref{fig:exactdiag} shows the PBE for two electrons added to a half-filled
$2\times 6$ ladder ($\Rightarrow n=7/6 \approx 1.17$) 
with $t_1 = 4/3$, $t_2=2/3$ ($\Rightarrow \Delta t = 2/3$), and $t_{\perp}=4/3$. Clearly, positive PBE are
observed only when $u_1$ and $u_2$ are close and $E^{-}_{pb}$ is the largest for intermediate values of $u_y$.

\begin{figure}
\includegraphics[width=0.40\textwidth]{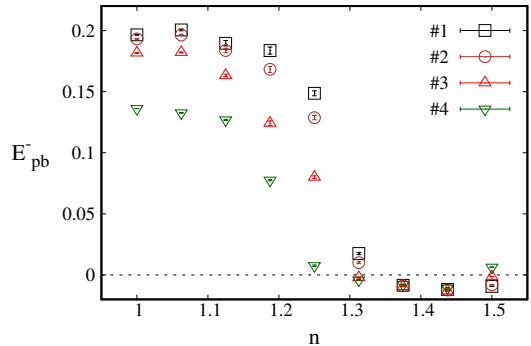}
\caption{\label{fig:U20EpbMinusVariousDopping} 
Pair binding energy $E^{-}_{pb}$ as function of the average density $n$ for ladders with $L=128$ rungs.
The model parameters are listed in Table~\ref{table:u1_Equal_u2}.
Error bars show the size of DMRG truncation errors.
}
\end{figure}

We also study the PBE as a function the density in the range $1 \leq n \leq 1.5$. 
For instance, Fig.~\ref{fig:U20EpbMinusVariousDopping} shows results for ladders with $L=128$ rungs.
The model parameters are the first four parameter sets listed in Table~\ref{table:u1_Equal_u2}
corresponding to  the maxima observed in Fig.~\ref{fig:U20EpbMinus}(a). We see that in all cases the PBE remain
clearly positive over a rather broad range of densities. Thus the phase diagram of the model~(\ref{eq:hamiltonian}) 
exhibits an extended phase with pair binding.

\begin{figure}[b]
\includegraphics[width=0.40\textwidth]{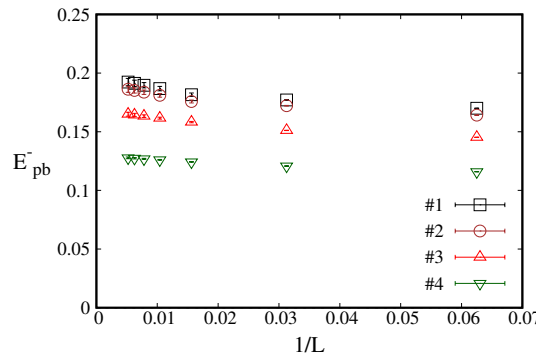}
\caption{\label{fig:EpbHolesExtrap} 
Pair binding energy $E^{-}_{pb}$ as function of the inverse ladder length for an average density $n=1.125$. 
The model parameters are listed in Table~\ref{table:u1_Equal_u2}.
Error bars show the size of DMRG truncation errors.
}
\end{figure}

\begin{figure}
\includegraphics[width=0.40\textwidth]{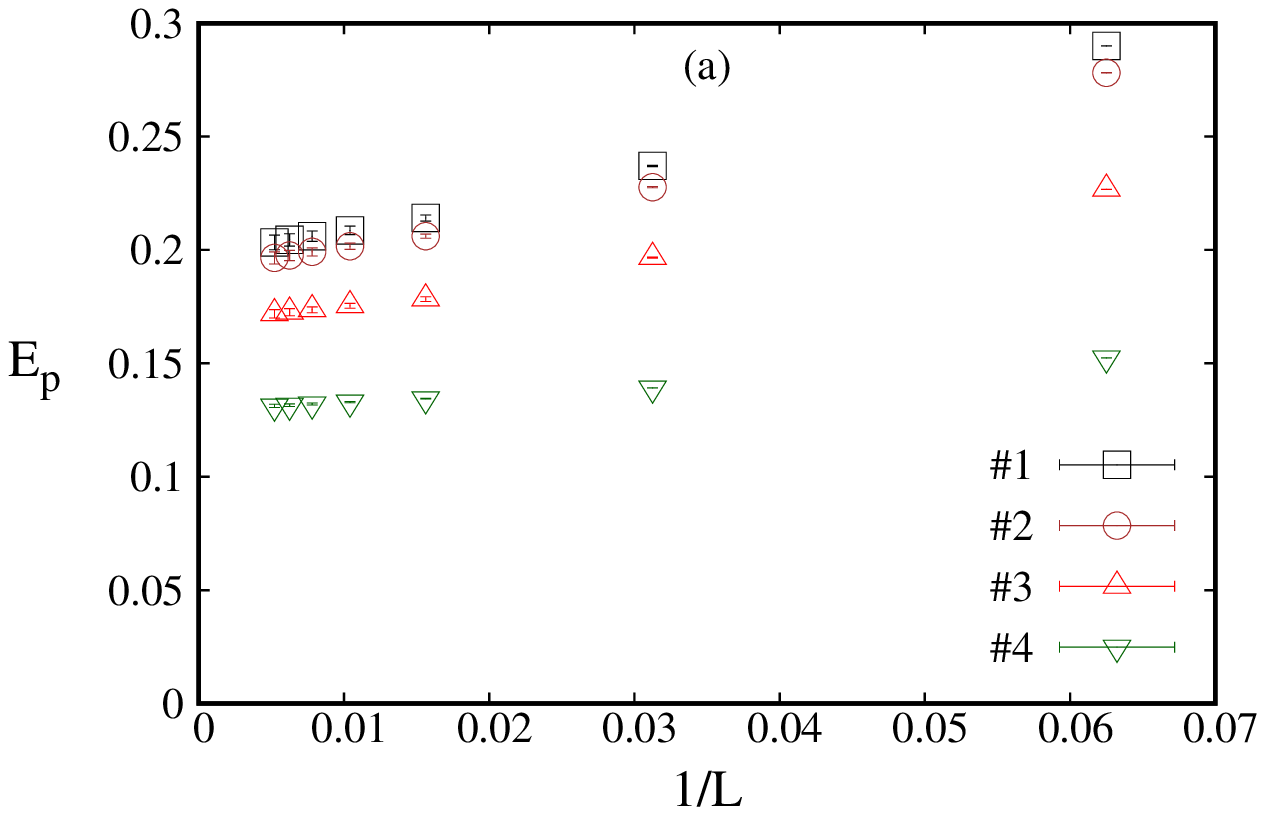}
\includegraphics[width=0.40\textwidth]{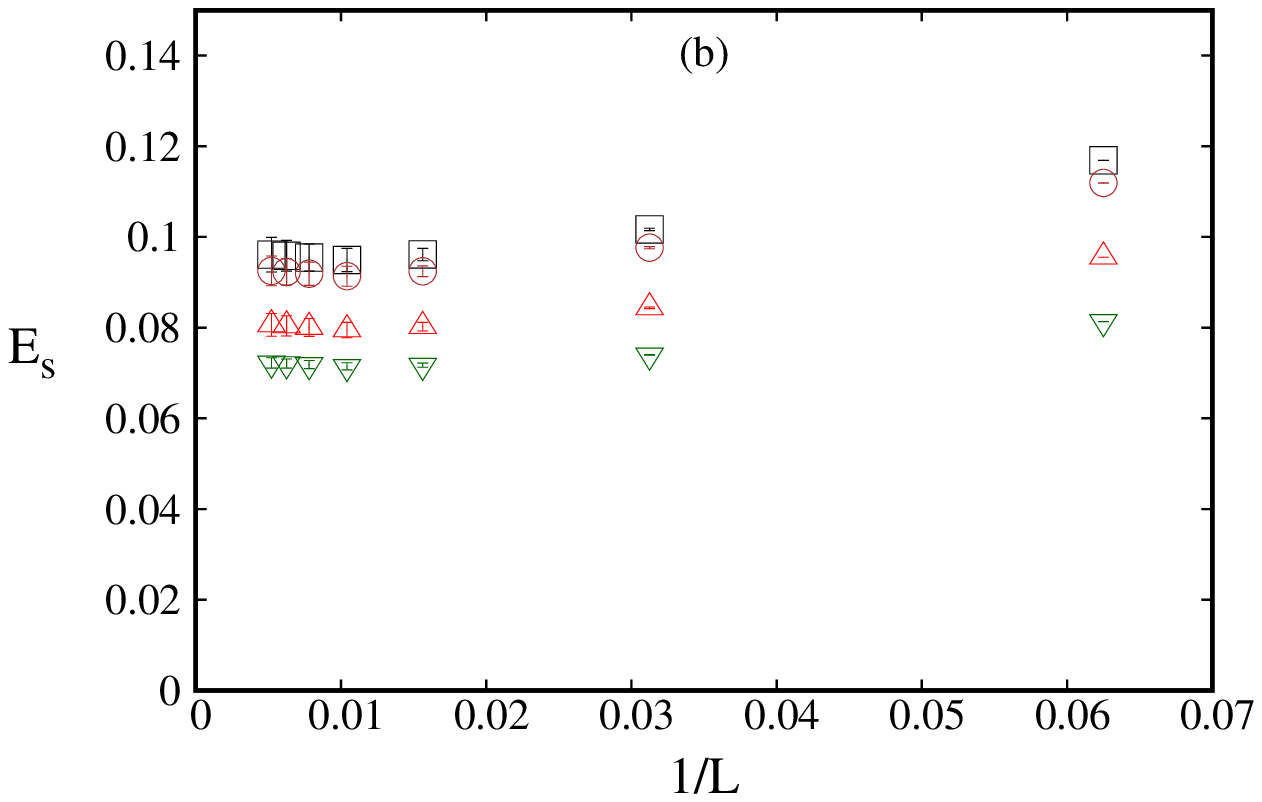}
\includegraphics[width=0.40\textwidth]{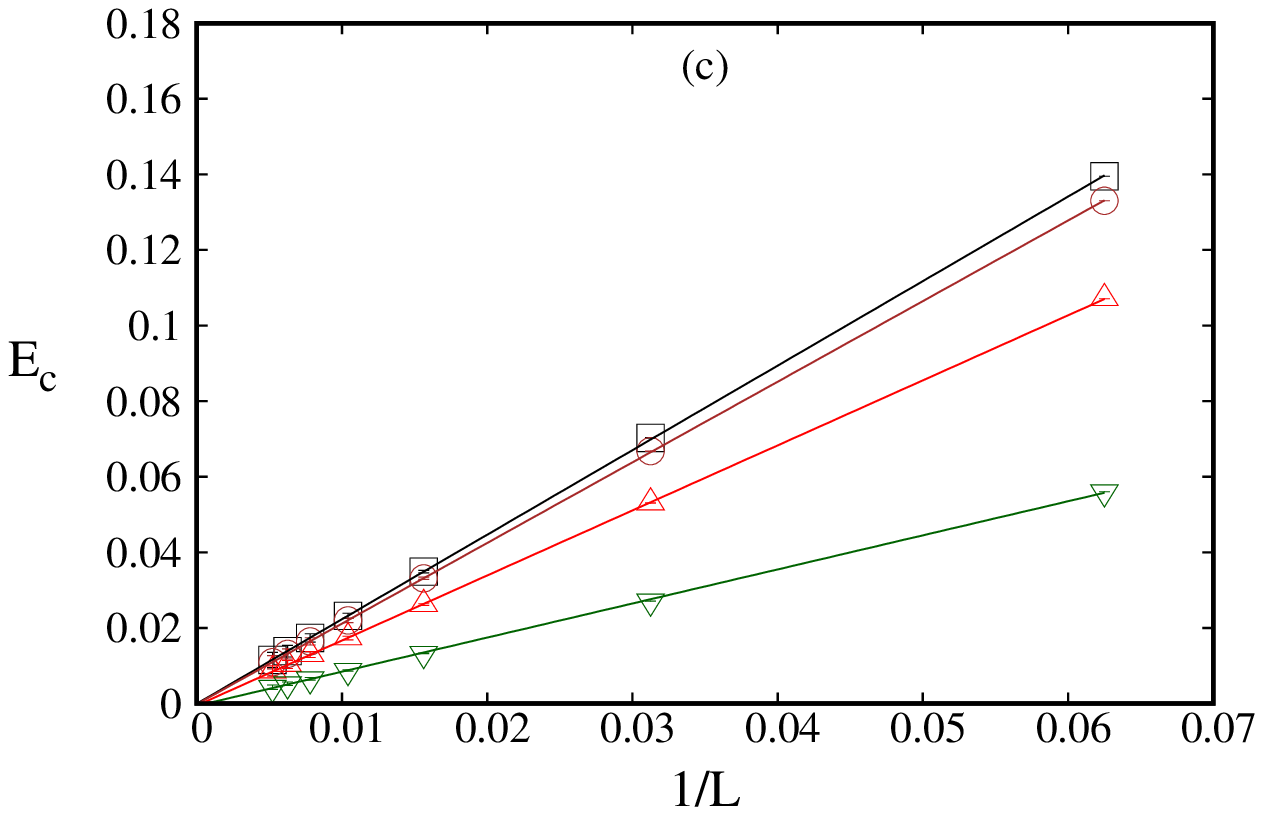}
\caption{\label{fig:EpEsEcExtrap} 
(a) Single-particle gap $E_p$, (b) spin gap $E_s$ and (c) charge gap $E_c$ as function of the inverse ladder length for an average density $n=1.125$.
The model parameters are listed in Table~\ref{table:u1_Equal_u2}.
Error bars show the size of DMRG truncation errors.
}
\end{figure}

To confirm the existence of pair binding in the thermodynamic limit we have analyzed $E^{-}_{pb}$  as a function of the ladder length $L$ up 
to $L=200$.  For instance, Fig.~\ref{fig:EpbHolesExtrap} shows  $E^{-}_{pb}$ as a function of $1/L$ for the first four parameter sets listed in
Table~\ref{table:u1_Equal_u2}, i.e. for the maxima observed in Fig.~\ref{fig:U20EpbMinus}(a).
We see clearly that the PBE approaches a finite value for $1/L\rightarrow 0$ in all four cases. 
Similarly, we find that the gaps for charge excitations vanish for long ladders while the gaps for spin excitations remains finite, see Fig.~\ref{fig:EpEsEcExtrap}.
These findings are compatible with a Luther-Emery phase~\cite{Luther1974}.

An important question is how energy is saved in the pair formation.
To answer this question we have analyzed the variations of the 5 contributions to the
ground state energies $E_0(N_{\uparrow},N_{\downarrow})$ and thus to the PBE separately:
the three hopping terms and the two interaction terms in~(\ref{eq:hamiltonian}).
Unfortunately, theses variations are significantly larger than $E^{-}_{pb}$
and almost cancel each other to result in the small positive or negative PBE reported here.
Thus it is not possible to determine what kind of energy is saved by forming the 
bound electron pairs.

\begin{figure}
\includegraphics[width=0.40\textwidth]{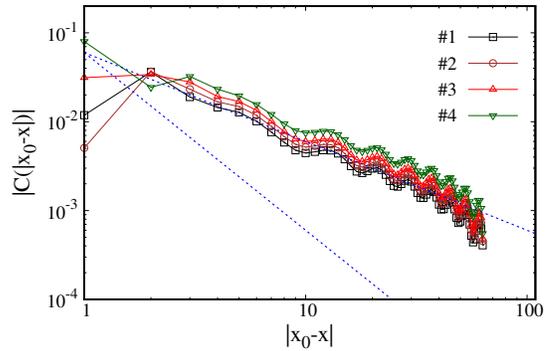}
\caption{\label{fig:PairPairCorrelationsRatio} 
Absolute value of the pair correlation function~(\ref{eq:RLcorrelation}) for ladders with $L=128$ rungs and a density $n=1.125$.
The dashed lines are $\sim \vert x_0 -x\vert^{-1}$ and $\sim \vert x_0-x\vert ^{-2}$.
The model parameters are listed in Table~\ref{table:u1_Equal_u2}.
}
\end{figure}

\begin{figure}[b]
\includegraphics[width=0.4\textwidth]{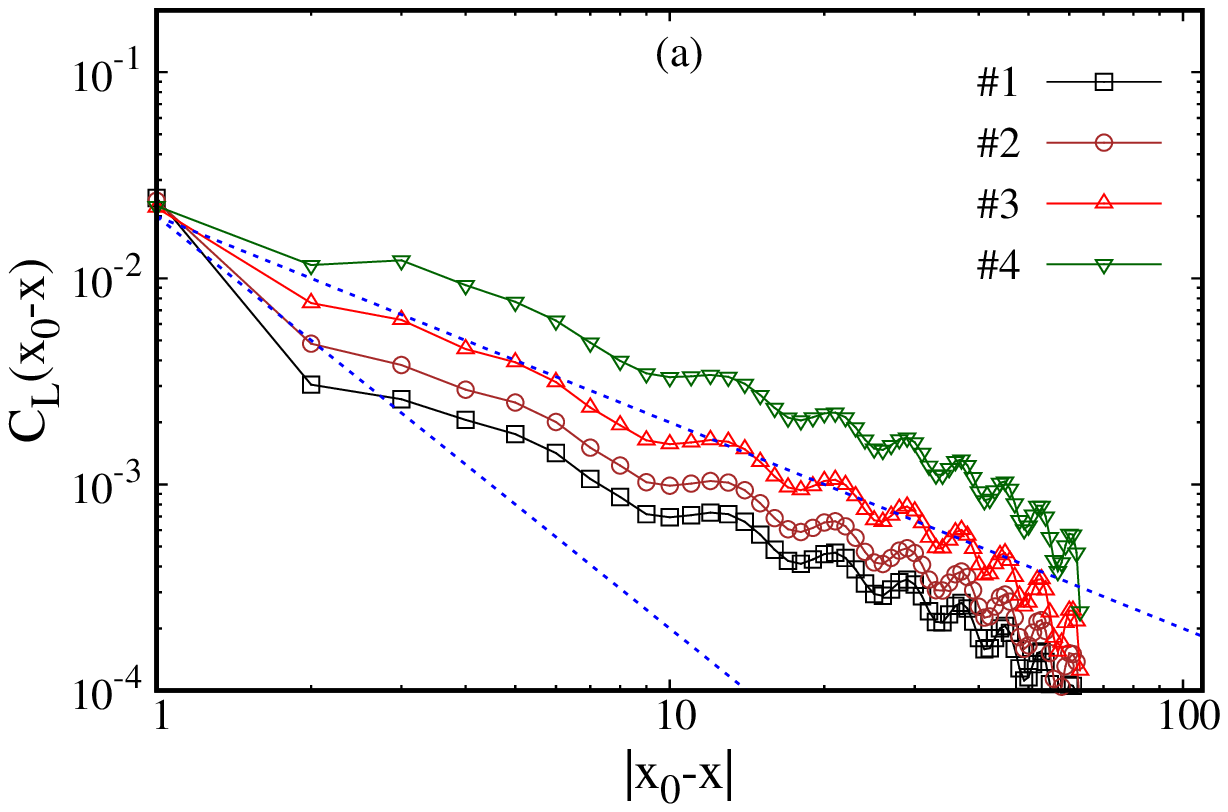}
\includegraphics[width=0.4\textwidth]{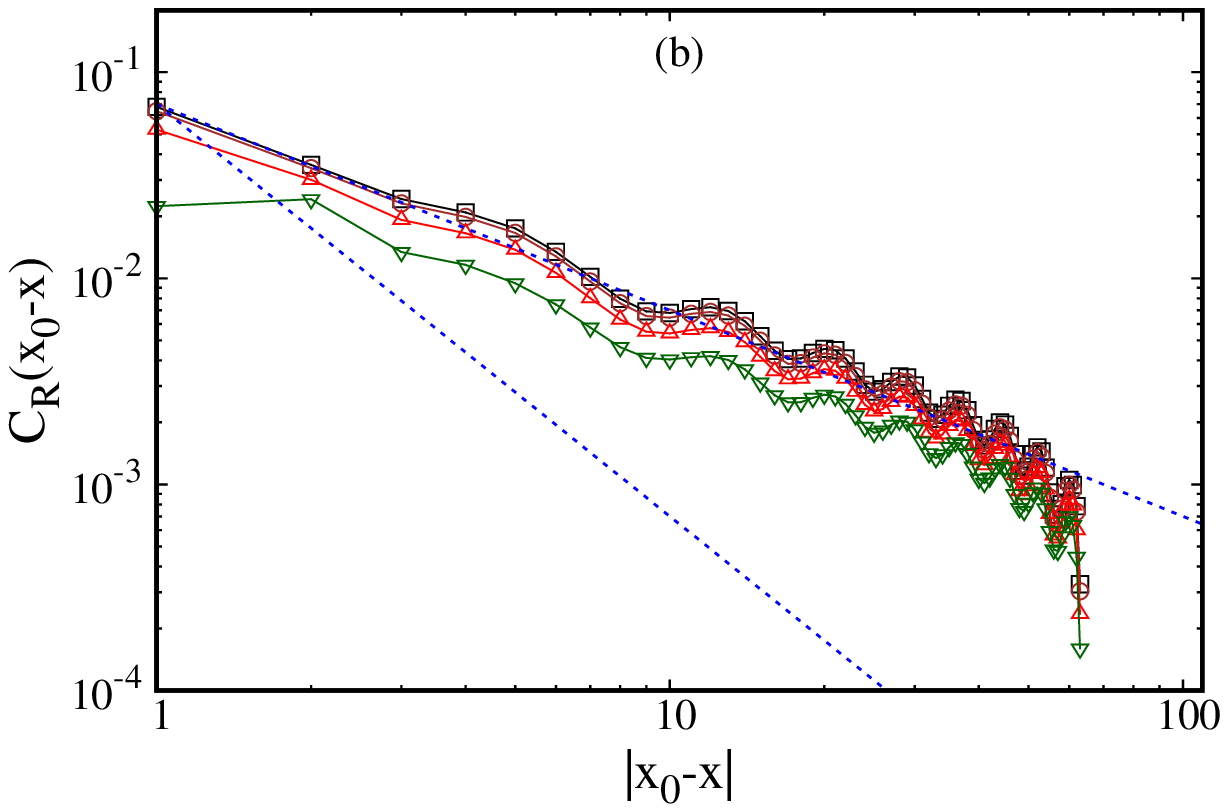}
\caption{\label{fig:CorrRatio} 
Absolute value of  pair correlation functions for ladders with $L=128$ rungs and a density $n=1.125$.
(a) For the leg-leg pair correlation function~(\ref{eq:LLcorrelation}) and (b) for the rung-rung pair correlation function~(\ref{eq:RRcorrelation}).
The dashed lines are $\sim \vert x_0 -x\vert^{-1}$ and $\sim \vert x_0-x\vert ^{-2}$.
The model parameters are listed in  Table~\ref{table:u1_Equal_u2}.
}
\end{figure}

\subsection{Pairing correlations}

Finally, we have studied pair correlation functions~(\ref{eq:correlations}) in finite-size ladders. 
Despite the difficulty of determining the long-range behavior of correlation functions
from finite-size systems, we observe clear evidence for the 
enhancement of pairing correlations where the PBE is positive. 
For instance, Fig.~\ref{fig:PairPairCorrelationsRatio} shows the rung-leg correlations~(\ref{eq:RLcorrelation})
for the  first four parameter sets listed in Table~\ref{table:u1_Equal_u2}
corresponding to  the maxima observed in Fig.~\ref{fig:U20EpbMinus}(a). 
Clearly,  they decay as $\vert x_0-x\vert^{-1}$. 
Other pair correlation functions, such as for rung-rung~(\ref{eq:RRcorrelation}) and leg-leg~(\ref{eq:LLcorrelation}) pairs, are also enhanced, as shown in Fig.~\ref{fig:CorrRatio}, but none decays
more slowly than the rung-leg correlations~(\ref{eq:RLcorrelation}).
We note that rung-leg correlations are negative, at least
for short distances $\vert x_0-x\vert$, while the rung-rung and leg-leg correlations are positive.
In symmetric ladders this is view as reminiscent of electron pairs with d-wave symmetry in 2D systems~\cite{noa97}. 
The fact that we find the same pairing structure in asymmetric ladders calls this analogy into question
because the systems studied here bear no relation to rotational symmetries in higher dimensions.

\begin{figure}
\includegraphics[width=0.40\textwidth]{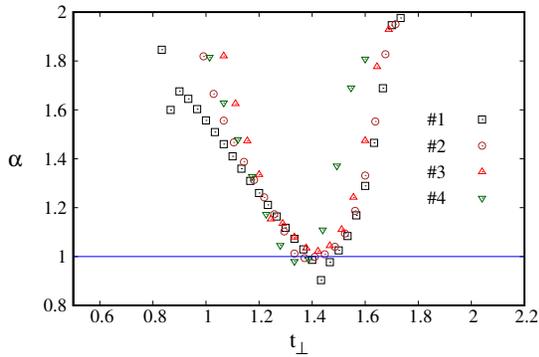}
\caption{\label{fig:ExponCorrRatio} 
Exponent $\alpha$ of pairing rung-leg correlations~(\ref{eq:RLcorrelation})
 as function of $t_{\perp}$ for ladders with $L=128$ rungs and a density $n=1.125$.
The model parameters are listed in  Table~\ref{table:u1_Equal_u2}.
}
\end{figure}

In particular,
we find the strongest quasi-long-range pairing correlations where the PBE reaches a maximum
as a function of $t_{\perp}$. To quantify the quasi-long-range order  we fit correlation functions 
 using a power law $\beta \vert x-x_0 \vert^{-\alpha}$ with fit parameters $\alpha$ and $\beta$.
 The resulting exponent $\alpha$  for the rung-leg correlation function~(\ref{eq:RLcorrelation}) 
  is shown in Fig.~\ref{fig:ExponCorrRatio}
for the  first four parameter sets listed in Table~\ref{table:u1_Equal_u2}.
Clearly, $\alpha$ is minimal and reaches the value 1 for $t_{\perp} \approx 1.4 t_{\parallel}$.

Finally, we have also analyzed density-density and spin-spin correlation functions.
We have not found
any significant qualitative difference between these correlations in symmetric and asymmetric ladders when
the PBE is positive and pairing correlations are enhanced. In particular, spin-spin correlations
decrease exponentially with distance in agreement with the observation of a small but finite
spin gap in that regime. Thus pairing, density, and spin correlations do not reveal
any difference in the pairing structure when the ladder asymmetry is modified.

\section{Discussion and outlook}
Our results show that the pairing occurring in symmetric Hubbard ladders  is robust against asymmetry in the legs as long as
$U_1/t_1 \approx U_2/t_2$. The pairing strength and the pair structure revealed by correlation functions remain similar
when an asymmetry is introduced. Thus it is questionable, whether the pairing found in symmetric ladders is a precursor
for (possible) superconductivity in the 2D one-band Hubbard model. In particular, the pairing structure observed in two-leg ladders
does not seem to be related to a possible d-wave symmetry in two dimensions. In our opinion, the ladder model~(\ref{eq:hamiltonian}),
including the symmetric case, is better seen as a 1D two-band correlated electron model than as a sublattice of higher dimensional system.

In contrast, a 2D system of weakly-coupled symmetric two-leg ladders seems to reproduce the d-wave pairing correlations of the Hubbard model on a square lattice~\cite{Maier2022}. It would be interesting
to investigate a 2D systems of coupled asymmetric
ladders to determine whether the pairing structure is
really related to a rotational symmetry of the square lattice.

An advantage of the model~(\ref{eq:hamiltonian}) is that we 
can study the influence of various interactions on the occurrence and strength of pairing. 
Numerically we have found pairing only at intermediate couplings (e.g. $u_y \approx 5$ to $10$ and $t_{\perp}/t_{\parallel} \approx 0.5$ to $1.5$).
For this reason, we could not yet  obtain an approximate analytical solution for the relevant parameter regime
that could explain the pairing mechanism. 
In particular, pairing does not occur in the strong interchain coupling regime (i.e., large $t_{\perp}$)
where strong rung singlets are formed. Thus the occurrence and structure of pairs cannot be explained
by the formation of rung (hole or doublon) pairs to preserve rung singlets on the other rungs.

Nevertheless, it is clear that when pairing occurs upon doping away from half-filling, the half-filled ladder is in
a crossover regime between a Mott insulator and a two-band charge transfer insulator.
On the one-hand, in the weak interchain coupling limit ($t_{\perp} \ll \vert t_y\vert$) 
the asymmetric ladder corresponds to two almost independent Hubbard chains,
which are known exactly to be Mott insulators~\cite{Lieb1968,Essler2001,Gebhard1997} at half filling. 
Low-energy charge excitations move along the legs in this limit.
On the other hand, in the strong interchain coupling  limit
($t_{\perp} \gg \vert t_y\vert$) the asymmetric ladder corresponds to a system of $L$
almost independent rungs. At half-filling its low-energy physics resemble a charge transfer insulator with localized excitations on the rungs.
(In the classification of Ref.~\cite{Zaanen1985}  charges transfer from
the bonding to the antibonding rung states separated by an energy gap $\Delta \approx 2t_{\perp}$.)
However, pairing occurs in the intermediate regime $t_{\perp} \approx  t_{\parallel}$, where charge excitations involve both
rung and leg directions.

Similarly, the physics of undoped cuprate compounds lies in between 
pure 2D charge transfer and Mott insulators.
On the one hand, they are often classified as 
charge transfer insulators  because it is easier for 
charge excitations to move between orbitals in the same unit cell than to move
between unit cells~\cite{Zaanen1985}. 
This physics can be described by a limit of the  three-band Hubbard model.
On the other hand, assuming that doped holes occupy primarily oxygen atoms,
the low-energy physics of the three-band Hubbard model at low doping can
be described by the one-band Hubbard model away from half filling~\cite{zha88}, which is a doped Mott insulator~\cite{Gebhard1997}.
However, realistic parameters for cuprate compounds lie in between these two limiting cases~\cite{Dagotto1994}.
Therefore, we think that further studies of the asymmetric ladder model~(\ref{eq:hamiltonian})
could provide us with useful information about the pair binding physics in correlated 
electron systems  between charge transfer and Mott insulators, like the
superconducting cuprate compounds.

\acknowledgments
The DMRG calculations were carried out on the cluster system at the Leibniz University of Hannover.


\end{document}